\documentclass[aps,pra,twocolumn,superscriptaddress,floatfix]{revtex4}

\pdfoutput=1

\usepackage{hyperref}

\usepackage{xspace}

\newcommand{\alphaD}{\ensuremath{\alpha_\mathrm{d}}\xspace}
\newcommand{\nbarSig}{\ensuremath{\bar{n}_\mathrm{sig}}\xspace}
\newcommand{\nbarBg}{\ensuremath{\bar{n}_\mathrm{bg}}\xspace}

\newcommand{\dd}{\mathrm{d}}

\newcommand{\ii}{\mathrm{i}}

\expandafter\let\csname equation*\endcsname\relax 			
\expandafter\let\csname endequation*\endcsname\relax

\usepackage{amsmath}
\usepackage{amsfonts}
\usepackage{amssymb}
\usepackage{graphicx}        

\newcommand{\bra}[1]{\ensuremath{\langle #1 \vert}\xspace}
\newcommand{\ket}[1]{\ensuremath{\vert #1 \rangle}\xspace}

\newcommand{\ketbra}[2]{\ket{#1}\bra{#2}}

\newcommand{\LG}[1]{\ensuremath{\mathrm{LG}^{l=#1}}\xspace} 
\newcommand{\OAM}[1]{\ensuremath{\ket{l=#1}}\xspace} 
\newcommand{\Right}{\ensuremath{\ket{\mathrm{R}}}\xspace} 
\newcommand{\Left}{\ensuremath{\ket{\mathrm{L}}}\xspace} 
\newcommand{\G}{\ensuremath{{\mathrm{TEM}_{00}}}\xspace} 
\newcommand{\HG}[1]{\ensuremath{\mathrm{HG}^{#1}}\xspace} 
\newcommand{\ketV}{\ensuremath{\ket{\mathrm{V}}}\xspace} 
\newcommand{\ketD}{\ensuremath{\ket{\mathrm{D}}}\xspace} 
\newcommand{\ketH}{\ensuremath{\ket{\mathrm{H}}}\xspace} 
\newcommand{\ketA}{\ensuremath{\ket{\mathrm{A}}}\xspace} 

\usepackage{numprint} 
\usepackage[cdot,thickspace,amssymb]{SIunits} 
\usepackage{xfrac}
\usepackage{array} 
\usepackage{multirow} 

\begin{document}

\title{Quantum state tomography of orbital angular momentum photonic qubits \\via a projection-based technique}

\author{Adrien Nicolas, Lucile Veissier, Elisabeth Giacobino, Dominik Maxein, and Julien Laurat}

\affiliation{Laboratoire Kastler Brossel, Universit\'{e}
Pierre et Marie Curie, Ecole Normale Sup\'{e}rieure, CNRS, Coll\`ege de France, Case 74, 4 place
Jussieu, 75252 Paris Cedex 05, France}

\begin{abstract}
While measuring the orbital angular momentum state of bright light beams can be performed using imaging techniques, a full characterization at the single-photon level is challenging. For applications to quantum optics and quantum information science, such characterization is an essential capability. Here, we present a setup to perform the quantum state tomography of photonic qubits encoded in this degree of freedom. The method is based on a projective technique using spatial mode projection via fork holograms and single-mode fibers inserted into an interferometer. The alignment and calibration of the device is detailed as well as the measurement sequence to reconstruct the associated density matrix. Possible extensions to higher-dimensional spaces are discussed.
\end{abstract}

\maketitle

\section{Introduction}
In the past decades, light beams carrying orbital angular momentum (OAM) have raised a considerable interest \cite{Padgett_03,Torres_11}. Laguerre-Gaussian beams for instance have been used in a variety of fields for applied and fundamental investigations, including hyperdense data transfer \cite{Wang_12}, particule-motion control \cite{Greier_03,Phillips_06}, enhanced-sensitivity measurements \cite{Patarroyo_13,Dambrosio_13}, fundamental tests of quantum mechanics \cite{Mair_01,Dada_11} or quantum information protocols \cite{Groeblacher_06, Simon_13, Fickler_14, Dambrosio_12,Krenn_14}. Quantum information science and technology has indeed recognized single photons in OAM state superpositions as promising information carriers as the high dimensionality of the Hilbert space they live in could enable enhanced security and multiplexing \cite{Bourennane_02}. In this framework, optical and quantum memories for OAM states have recently seen tremendous developments \cite{Inoue_09,Pugatch_07,Moretti_09,Veissier_13,Ding_13,Nicolas_14}, opening the path to quantum networks and scalable communication architectures based on this degree of freedom. 

In these quantum information related applications, it is crucial to measure the orbital angular momentum state of single photons, i.e. to perform the full quantum state tomography. While OAM measurement for bright beams of light can be done via imaging, it is much more challenging in the single-photon regime. The various solutions that have been proposed and tested heretofore can be classified into two categories. The earlier methods perform single-photon detection after the signal has passed through a projector on an OAM eigenstate. They typically have low efficiencies but good selectivity. The more recently-developed mode-sorting techniques consist in changing the signal path depending on its OAM value. These methods usually lead to higher efficiencies but their implementation is also often more challenging.

In this paper, we describe an interferometer-based detection setup relying on the projective method, similar to the one that Miyamoto and coworkers used in order to overcome the limitations of the ``hologram shifting method'' \cite{Miyamoto_11}, but with an improved efficiency and versatility. 
Depending on the interferometer phase, our device allows to perform state projection into multiple bases and consequently a full state tomography. This technique has been used in our recent demonstration of quantum memory for OAM encoded qubits \cite{Nicolas_14}. We provide here a detailed and quantitative study of its implementation, and investigate extensions to higher-dimensional spaces.

The paper is organized as follows. Section~\ref{sec:qubits}  first briefly reminds the fundamentals on quantum bits and on orbital angular momentum in order to understand our implementation and the required measurements. The full description of our tomography setup is then given in section \ref{sec:device}. In section \ref{sec:calib}, a detailed study of the calibration procedure is presented and benchmarks are given. In section \ref{sec:data}, we finally show examples of qubit state reconstruction. Perspectives are discussed in section~\ref{sec:perspectives} where we propose two possible extensions of the setup to perform on higher-dimensional OAM Hilbert spaces.

\section{Qubits encoded in the orbital angular momentum degree of freedom}
\label{sec:qubits}

In this section, we first summarize the representation of qubit states and the required measurements to reconstruct the associated density matrix. We then discuss specifically the implementation of OAM qubits with superpositions of Laguerre-Gaussian modes. 

\subsection{Qubit representation and quantum state tomography}
A qubit, i.e. a two-dimensional quantum system, evolves in a Hilbert space spanned by two basis vectors usually denoted $\ket{0}$ and $\ket{1}$ in analogy with classical information. For pure state, it can be represented by:
\begin{equation}
\label{Eq:Qubit}
\ket{\Psi} = \alpha \ket{0} + \beta \ket{1}
\end{equation}
with $|\alpha|^2+|\beta|^2 = 1$.

The tomography  of such superposition states requires measurements performed in three different bases \cite{James_2001}. In addition to the logical basis $\{\ket{0}$,$\ket{1}\}$, two additional mutually unbiased bases can be defined as the superpositions: $\{\ket{\pm}=\ket{0}\pm\ket{1}\}$ and $\{\ket{\pm\ii}=\ket{0}\pm \ii \ket{1}\}$. Measuring the qubit in the logical basis will yield the values of $|\alpha|^2$ and $|\beta|^2$ and measurements in the two superposition bases will provide the relative phase between the coefficients $\alpha$ and $\beta$. This sequence of measurement actually allows to perform the reconstruction of the density matrix of any mixed state, which in the general case can be expressed as :
\begin{equation}
\label{Eq:density_matrix}
\hat{\rho}  = \frac{1}{2} \left( \hat{\mathbb{I}} + \sum_{i=1}^3 S_i \hat{\sigma}_i \right)
= \frac{1}{2} \left( \begin{matrix}
1+S_1 & S_2-i\,S_3 \\ 
S_2+i\,S_3 & 1-S_1
\end{matrix} \right)\;\;\textrm{,}
\end{equation}
where the $\hat{\sigma}_i$ are the Pauli matrices and the $S_i=\textrm{Tr}{\left(\hat{\rho}\hat{\sigma}_i \right)}$ coefficients are usually called Stokes parameters when dealing with polarization states. The $S_i$ coefficients indicate the relative weights of either basis state in the different bases. Indeed, they can written as $S_1 = p_{\ket{0}} - p_{\ket{1}}$, $S_2 = p_{\ket{+}} - p_{\ket{-}}$ and $S_3 = p_{\ket{+\ii}} - p_{\ket{-\ii}}$, where $p_{\ket{\Psi}}$ is the probability to measure the qubit in the state $\ket{\Psi}$. If the qubit is in a pure state as expressed in equation \ref{Eq:Qubit}, then \begin{equation}
\nonumber
\hat{\rho} = \ketbra{\Psi}{\Psi} = \left|\alpha\right|^2\ketbra{0}{0}+\left|\beta\right|^2\ketbra{1}{1}+\alpha\beta^*\ketbra{0}{1}+\beta\alpha^*\ketbra{1}{0}\;,
\end{equation} 
and the Stokes parameters can be easily related to the parameters $\alpha$ and $\beta$:
\begin{equation}
\nonumber
S_1 = |\alpha|^2-|\beta|^2 \qquad S_2 = 2Re(\alpha\beta^*) \qquad S_3 = -2Im(\alpha\beta^*).
\end{equation}

In this work, we focus on qubits encoded in the orbital angular momentum degree of freedom. We now briefly review the essential features of the orbital angular momentum of light, focusing on the Laguerre-Gaussian modes that offer a convenient basis to describe it. 

\subsection{Properties of the Laguerre-Gaussian modes}
The set of Laguerre-Gaussian modes is a complete orthonormal basis of solutions to the paraxial propagation equation. Their transverse shape is propagation invariant, as can be seen from the following expression of their electric field amplitude:
\begin{eqnarray}
\label{Eq_LG}
\mathrm{LG}^{l}_{p} (r, \theta, z) &=& \mathcal{E}_0 \frac{K_{lp}}{w(z)} \left( \frac{r\sqrt{2}}{w(z)} \right)^{\vert l \vert} e^{\ii l \theta} e^{-(\sfrac{r}{w(z)})^2} \\
&& \mathrm{L}^{\vert l \vert}_p \left(\frac{2r^2}{w(z)} \right) e^{-\ii k \sfrac{r^2}{2 R(z)}} e^{\ii(2p + \vert l \vert +1)\zeta(z)}\nonumber
\end{eqnarray}
where $\mathcal{E}_0$ is the electric field amplitude, $K_{lp} = \sqrt{\frac{2}{\pi} \frac{p!}{(l+p)!}}$ is a normalization constant, and $\mathrm{L}^{\alpha}_n (x)$ are the generalized Laguerre polynomials. Here, we have chosen a wave traveling along the $z$ axis and polar coordinates (r, $\theta$) to parametrize the transverse plane. The parameters $w(z)=w_0\sqrt{1+(\sfrac{z}{z_R})^2}$, $z_R=\pi \sfrac{w_0^2}{\lambda}$, $R(z) = z \left( 1 + (\sfrac{z_R}{z})^2 \right)$, and $\zeta(z) = \mathrm{arctg}(\sfrac{z}{z_R})$ are the radius, Rayleigh length, radius of curvature and Gouy phase for a beam of waist $w_0$ at a wavelength $\lambda$.

In contrast to the standard $\G = \mathrm{LG}^{l=0}_{p=0}$ mode, the higher-order Laguerre-Gaussian modes have a rotating phase profile $e^{\ii l \theta}$ with a singularity at the origin. Due to this rotating phase, the local Poynting vector has a non-vanishing component along the orthoradial direction and the beam thus exhibits an orbital angular momentum around this axis. The orbital angular momentum carried by each photon in such a mode is equal to the index $l \in \mathbb{Z}$ (in $\hbar$ units), which is also equal to the circulation of the phase around the axis divided by $ 2 \pi$. As the LG modes are eigenfunctions of the propagation equation, their transverse shape, and hence the OAM of the photons, is preserved. This makes the OAM number $l$ a relevant quantum number for information encoding. As required for the smoothness of the electric field amplitude, the phase singularity is associated with an intensity nulling in the middle as the $2l$-th power of the radial coordinate $(\sfrac{r}{w(z)})^{2 \vert l \vert}$. This feature gives the LG modes their characteristic doughnut-shaped intensity profiles.

\begin{figure}
\includegraphics[width=0.95\columnwidth]{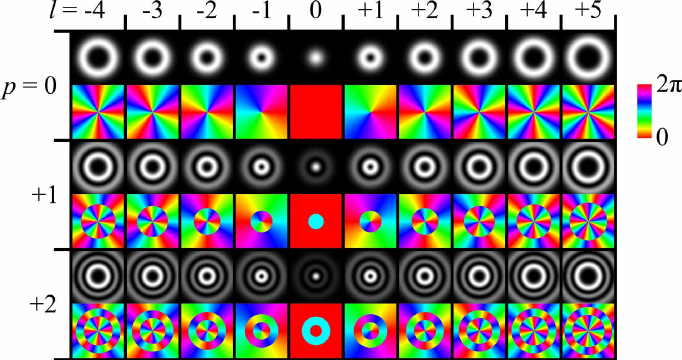}
\caption{Intensity and phase profiles of the first modes of the Laguerre-Gaussian family. Azimuthal index $l$ increases from left to right and radial index $p$ increases from top to bottom.}
\end{figure}

The other index, $p \in \mathbb{N}$, describes the radial shape of the beam. With each additional $p$ unit, the amplitude has an additional sign change along the radius and the intensity an additional zero-value ring. This number has a less straightforward interpretation than the orbital index $l$ and has therefore been subject to less investigation hitherto \cite{Plick_13,Karimi_14}.

\subsection{Superpositions of Laguerre Gaussian modes and OAM qubit encoding}
\label{SubSec:Qubit_OAM}

One promise of the OAM degree of freedom for information encoding is the large (potentially infinite) dimension of the available Hilbert space. We first restrict ourselves to a two-dimensional space spanned by the modes $\mathrm{LG}^{l=\pm 1}_{p=0}$. Extensions of the tomography scheme to higher-dimensional OAM spaces are discussed in section \ref{sec:perspectives}. 

For a single photon living in this qubit space, we define the corresponding logical basis vectors as $\ket{0}=\OAM{+1}=\Right$ and $\ket{1} = \OAM{-1} = \Left$. $R$ and $L$ respectively refer to the right and left handedness of the helical wavefront. The OAM difference of $2$ between the modes defining the logical basis ensures a very good distinguishability as will be explained later. As the index $p$ does not play a significant role in the present study, we drop it and we consider only modes with radial index $p=0$.

\begin{figure}
\begin{center}
\includegraphics[width=0.7\columnwidth]{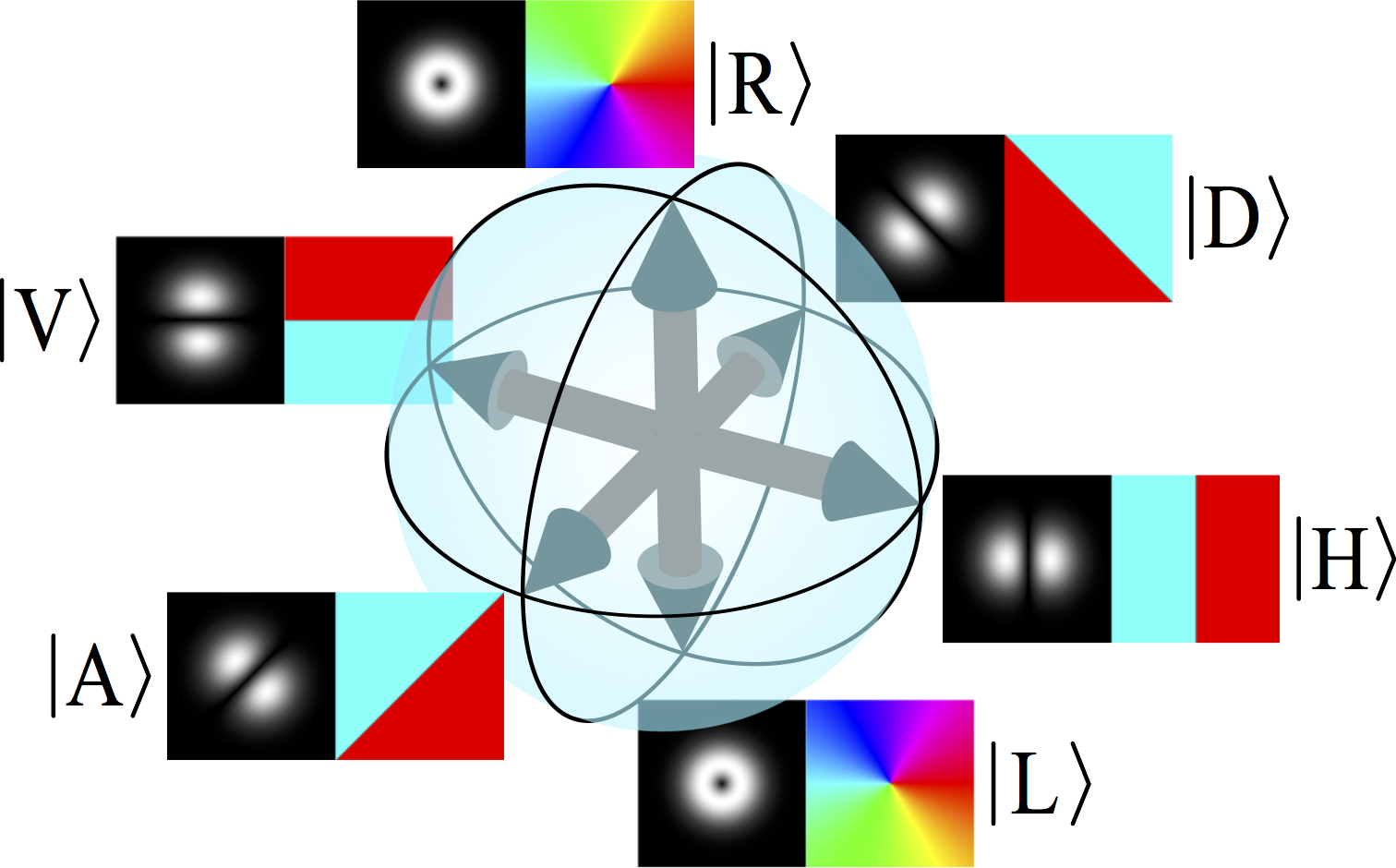}
\end{center}
\caption{Bloch sphere for the qubit basis $\{\Right$, $\Left\}$ and representation of the intensity and phase profiles of some relevant basis modes.}
\label{Fig:Bloch_sphere}
\end{figure}

Equally-weighted superpositions $\ket{\Psi} = \Right + e^{i \phi} \Left $, which span the equatorial plane of the Bloch sphere as shown in Figure~\ref{Fig:Bloch_sphere}, correspond to rotated Hermite-Gaussian (HG) modes of the type $\mathrm{TEM}_{01}$. These modes consist of two bright spots, separated by a dark line. In the plane orthogonal to the propagation axis, the angle \alphaD of the dark axis with respect to the horizontal axis is related to the relative phase $\phi$ by:
\begin{equation}
\label{Eq:alpha_dark_line}
\alphaD = \sfrac{(\phi-\pi)}{2}\;.
\end{equation}
Accordingly to their spatial shapes (see Fig.\ \ref{Fig:Bloch_sphere})  and in analogy to the case of a polarization basis, we denote the specific following modes as vertical, diagonal, horizontal and anti-diagonal:
\begin{equation}
\label{Eq:HG_modes}
\begin{split}
\begin{split}
\ketV & = (\Right - \phantom{\ii}\Left ) / \sqrt{2}\;,\\
\ketA & = (\Right - 		      \ii \Left ) / \sqrt{2}\;,
\end{split}
\;\;\;\;\;\;\;\;
\begin{split}
\ketD & = (\Right + 			\ii \Left )/ \sqrt{2}\;,\\
\ketH & = (\Right + \phantom{\ii}\Left ) / \sqrt{2}\;.
\end{split}
\end{split}
\end{equation}
The bases $\left\{\Right, \Left\right\}$, $\left\{\ketH, \ketV\right\}$ and $\left\{\ketD, \ketA\right\}$ will constitute the 3 mutually unbiased bases for the tomography.

\subsection{Detection of single photons in Laguerre-Gaussian modes}

The determination of the OAM state of a bright beam of light can be done by standard imaging and wavefront measurements (either using interferometry or a microlens array, or any equivalent available technique). This intrinsically requires many photons, so that characterizing an OAM state at the single-photon level calls for other methods. The different techniques developed so far can be classified into two categories:

\begin{itemize}
\item{\textit{Projective-based techniques.} In these methods, the photons to be measured impinge on a device that performs a projection onto an OAM eigenstate before being measured. The mode projectors are typically made of a hologram that converts an input mode with non-zero $l$ value into a \G mode followed by a spatial filter (pinhole or single-mode fiber). The holograms can be either fixed\cite{Mair_01} or dynamically programmed with a spatial light modulator \cite{Dada_11,Inoue_09}, they can be either intensity \cite{Vasnetsov_01} or phase holograms\cite{Mair_01}, they can diffract the light to all directions or be optimized for a single output direction. In any case, the projector only selects one mode and photons in other modes are lost.}
\item{\textit{Mode-sorting techniques.} Here, the propagation direction of the signal is changed depending on its OAM value. This feature overcomes the problem of losses inherent to mode projection. However, these methods are often more challenging than the previous ones. Among them, one can cite a Mach-Zehnder interferometer in the arms of which Dove prisms have been inserted \cite{Leach_02}. Another method that has seen significant developments in the few past years relies on a log-polar coordinate interpolation realized with two phase-modulating elements. The radial and polar coordinates $r$, $\theta$ in one plane are mapped one by one onto the cartesian $x$ and $y$ coordinates in a subsequent plane. This approach was first implemented with two spatial light modulators \cite{Berkhout_10} then with fixed refractive optics \cite{Lavery_12,Sullivan_12,Mirhosseini_13}. It is also possible to take benefit from the fact that any unitary manipulation of transverse modes (and hence mode sorting) can be achieved by multiple phase modulation steps separated by optical Fourier transforms \cite{Morizur_10}.}
\end{itemize}

\begin{figure*}[t!]
\includegraphics[width=0.77\linewidth]{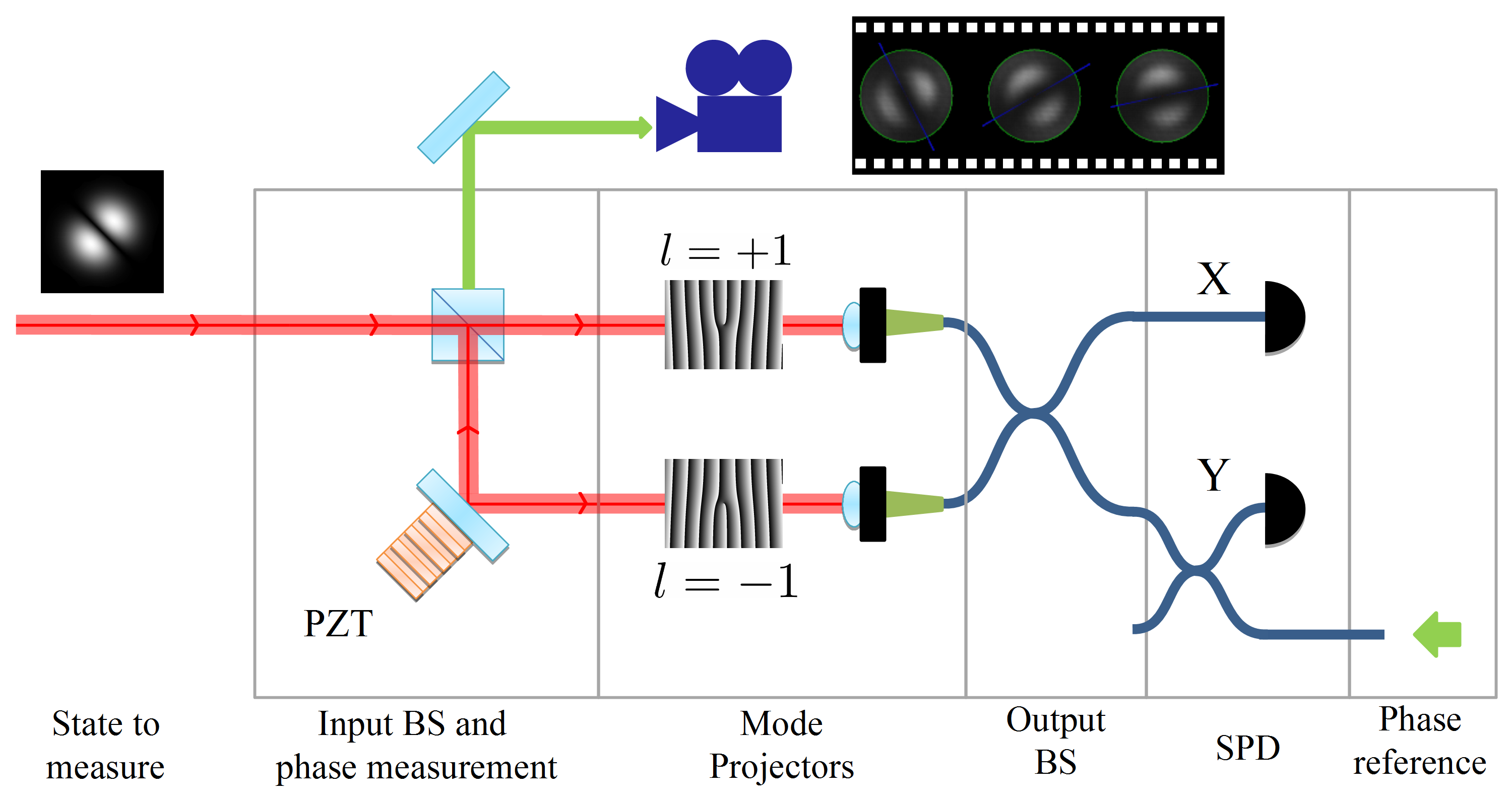}
\caption{Experimental setup for OAM qubit tomography. The state to be characterized enters a two-path interferometer, where each path includes a mode projector based on a blazed fork computer-generated hologram (phase pattern shown) and a single-mode fiber. Holograms labelled $l=+1$ and $l=-1$ are respectively placed in the so-called $R$ and $L$ paths. By OAM subtraction or addition, as defined by the orientation of the centered dislocation, they transform modes with orbital quantum number $l=+1$ (resp. $l=-1$) into \G modes, which are coupled into single-mode fibers performing spatial filtering. The fiber paths are then recombined and the two outputs are directed towards output $X$ and $Y$ where single-photon counting modules are located (SPCM-AQR-14-FC). A phase reference beam (green arrows) is injected backwards and detected by a digital camera at the input beam-splitter in order to measure the phase $\varphi$ of the interferometer. The value of $\varphi$ defines the projection basis.}
\label{Fig:Experimental_setup}
\end{figure*}

As underlined before, quantum state tomography requires projection into different bases in order to access the coherence terms of state superpositions. In this context, building on the projective techniques in order to benefit from their simplicity and high distinction ratio, we developed an OAM tomography setup, as described in the next section.

\section{Setup for quantum state tomography of OAM qubits}
\label{sec:device}

The state tomography requires to project the state to be characterized into three bases. In this section, we present an interferometric setup based on fork holograms enabling such projections. We explain how the interferometric phase defines the projection basis and we detail how to experimentally access this phase via an imaging technique.

\subsection{Experimental setup: Interferometer and mode projectors}

The apparatus is schematized in Figure~\ref{Fig:Experimental_setup}.
The incoming state is first split using a non-polarizing beam splitter. Each of the subsequent paths includes a mode projector onto an OAM eigenstate. These mode projectors are based on the combination of a hologram and a single-mode fiber \cite{Mair_01}.
A blazed fork phase-hologram diffracts the light and performs OAM addition or subtraction depending on its orientation.
Thus, on one path, the mode \LG{+1} is converted into a mode \LG{0}= \G, which is then efficiently coupled to the single-mode fiber, while any other mode is converted into a Laguerre-Gaussian beam with a non-vanishing $l$ value and hence not coupled to the subsequent fiber. 
There are two such paths, denoted $R$ and $L$, that are arranged to project the incoming state onto the \Right and \Left states respectively.
The diffraction efficiency of the holograms is $80\%$ and the coupling efficiency to the single-mode fiber is also around $80\%$, leading to an overall transmission around $65\%$ for the corresponding mode. The rejection of the other mode is optimized using the alignment described later, with typical value around \unit{25}{\deci\bel}.

As shown in Figure~\ref{Fig:Experimental_setup}, the two paths are then recombined via a fiber beam-splitter with two outputs labelled $X$ and $Y$. The difference in propagation length along the two arms of the interferometer causes a phase shift denoted $\varphi$.
If the input is in a Laguerre-Gaussian mode \Right or \Left, then light will only be coupled into one of the interferometer arms, and this light will be equally distributed over $X$ and $Y$ regardless of $\varphi$.
In contrast, if the input is in a superposition state, then there is a non-zero amplitude in both arms and these amplitudes will interfere.
The probabilities to detect light at either outputs $X$ or $Y$ will vary sinusoidally with $\varphi$.

More specifically, let us take the example of a pure state $\ket{\Psi} = a \Right + b e^{\ii \phi} \Left$ (with $a,b \in \mathbb{R}$) entering the device. At the output, it will be transformed into the state:
\begin{equation}
\label{Eq:output}
 (a+b e^{\ii (\phi+ \varphi)})\ket{X} + (a-b e^{\ii (\phi + \varphi)})\ket{Y}\; ,
\end{equation}
where we have assumed perfect transmission and mode rejection. The events detected at the output $X$ for instance will thus be proportional to 
\begin{equation}
 P=a^2+b^2+2ab\cos(\phi+\varphi).
\end{equation}
They correspond thus to the projection of the incoming state on the state $\Right + e^{\ii \varphi} \Left$. By choosing $\varphi$, any projection basis in the equatorial plane of the Bloch sphere can thus be chosen; this is the key feature of this setup.

In summary, the interferometer directs light in a given Hermite-Gaussian state towards output $X$ and the orthogonal mode towards output $Y$. These two states, i.e.\ the basis in which this separation occurs, depend on the value of the interferometer phase $\varphi$. For $\varphi = 0$, photons in the \ketH state will be directed towards output $X$ while photons in state \ketV will be directed towards output $Y$. In the same way, for $\varphi = \pi/2$, the incoming photons are measured with respect to the basis \ketA and \ketD. Finally, if one of the paths is blocked, then the device will act as a projector onto the OAM eigenstate \Right or \Left. In this case, the detectors at $X$ and $Y$ receive the same signal. Therefore, when the device is properly calibrated (as explained in section \ref{sec:calib}), it can be regarded as a black box performing state projection and yielding photon counts in the outputs $X$ and $Y$ as summarized in Table~\ref{Tab:detected_modes}. 

\begin{table}[t!]
\centering
\begin{tabular}{|c|c|c|c|c|c|c|}
\hline
& L path & R path & \multicolumn{4}{c|}{Full interferometer} \\
\cline{4-7} 
& blocked & blocked  & $\varphi=0$ & $\varphi=\pi$/2 & $\varphi=\pi$ & $\varphi=3\pi$/2 \\ 
\hline \hline 
Output X & \Right & \Left & $\ket{\rm{H}}$ & $\ket{\rm{A}}$ & $\ket{\rm{V}}$ & $\ket{\rm{D}}$ \\ 
\hline 
Output Y & \Right & \Left & $\ket{\rm{V}}$ & $\ket{\rm{D}}$ & $\ket{\rm{H}}$ & $\ket{\rm{A}}$ \\ 
\hline 
\end{tabular}
\caption{Detected projections in the output $X$ and $Y$ as a function of the configuration of the interferometer. The relative phase $\varphi$ defines the projection basis.}
\label{Tab:detected_modes}
\end{table}
 
\subsection{Scanning and measurement of the interferometer phase}

In order to change the projection basis, one mirror inside the interferometer is mounted on a piezoelectric transducer, allowing to vary $\varphi$. In the following, we explain how to access this phase using backwards-propagating reference light.

\subsubsection{Phase reference beam.}

\label{Subsec:phase_reference}
Thermal and mechanical drifts continuously change the interferometer phase $\varphi$ on a scale of a few degrees in a few seconds. 
To access this phase, a phase-reference beam is injected backwards into the interferometer. 
While it propagates backwards, the reference beam crosses the two holograms. The \G modes emerging out of the fibers are thus converted into an \LG{-1} mode in the $R$ path, and into an \LG{+1} mode in the $L$ path. 
These modes are superimposed at the input beam-splitter with a phase difference that is precisely equal to the interferometer phase $\varphi$.
As a result, we get the superposition $\left(\LG{+1} +e^{i \varphi} \LG{-1}\right)$. Similarly to equations \ref{Eq:alpha_dark_line} and \ref{Eq:HG_modes}, the equal weight superposition of $\LG{+1}$ and $\LG{-1}$ with a relative phase $\varphi$ results in a rotated Hermite-Gaussian $\mathrm{TEM}_{01}$ mode, consisting of two bright spots with a dark line between them. The dark line makes an angle 
\begin{equation}
\label{eq:alphaDInterferometer}
\alphaD = (\varphi-\pi)/2
\end{equation}
with the horizontal axis. Measuring this angle enables us to access the interferometer phase $\varphi$. We now detail how to analyze the images taken by the digital camera to efficiently extract this information.

\begin{table}[b]
\begin{center}
\begin{tabular}{|c|c|c|}
\hline
Image & Contrast & \\
without & enhancements & Fitted center \\
enhancement & \& projections & \\
\hline
\multirow{5}{*}{\raisebox{-1.55ex}{\includegraphics[scale=0.12]{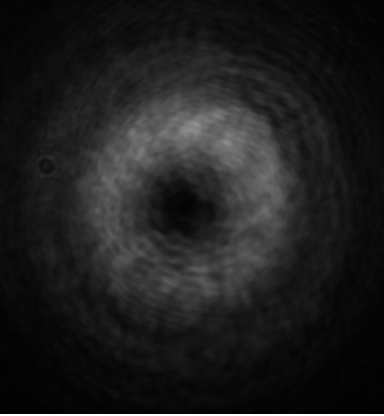}}} & \multirow{5}{*}{\raisebox{-1.55ex}{\includegraphics[scale=0.12]{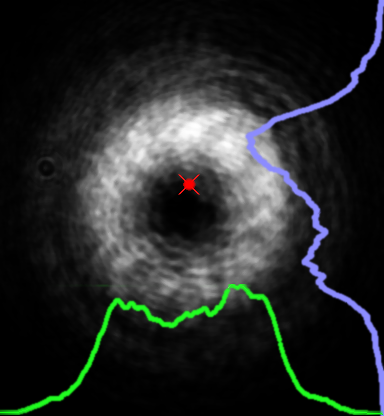}}} & \multirow{5}{*}{\raisebox{-1.55ex}{\includegraphics[scale=0.12]{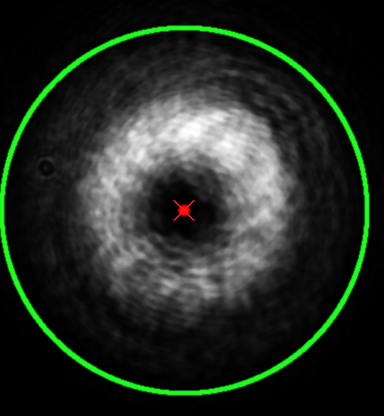}}} \\
& & \\
& & \\
& & \\
& & \\
\hline
\end{tabular}
\caption{\label{Tab:ImageAnalysis_center}Illustration of the steps involved in determining the center of symmetry of the images. The image displayed in the first column is the average of a few hundred phase-reference pictures. After treating it with the median filtering and midtone stretching operations explained in the text, the second image is obtained. Their difference shows the importance of the image processing; without it, the background would be strong enough to shift the center out of its actual position. In the second image, the projections of the intensity onto $x$ and $y$ axes are shown. They are used to obtain the starting values for the fit: the first-order momenta give an estimated center position (red dot), the second-order momenta a starting value for the width of the ring. The last column shows the fit output: the center is depicted in red and the circle with the radius of interest in green.
}
\end{center}
\end{table}

\subsubsection{Image analysis routine.}
\label{pythonroutine}

In a first step, all images are enhanced by applying a median filter (to reduce high-frequency noise and dead pixels) and a midtone stretching filter (to increase the contrast in the middle intensity region and to reduce variations in the high- and low-intensity regions).

When $\varphi$ varies, the dark line in the intensity pattern of the reference beam rotates around the beam axis, according to equation \ref{eq:alphaDInterferometer}. First, the center around which the dark line rotates has to be determined.
For this purpose, many images for different values of $\varphi$ are required, covering roughly uniformly the whole range of $\varphi \in [0,2\pi[$ (corresponding to the range of $[0,\pi[$ for the angle $\alphaD$ of the dark line).
Averaging them results in a ring-shaped image, as shown in Table~\ref{Tab:ImageAnalysis_center}, to which a doughnut-like distribution is fitted. The fit provides two parameters: the position of the center and the radius of interest.
As long as the alignment of the reference path is not changed, these values remain valid for all images. Consequently, this initial procedure has to be performed only once for a measurement series: either during the calibration, if a real-time analysis of the phase is desired, or using a part of the stored images, if post-processing is performed.

\begin{figure}[t!]
\centering
\includegraphics[width=0.95\columnwidth]{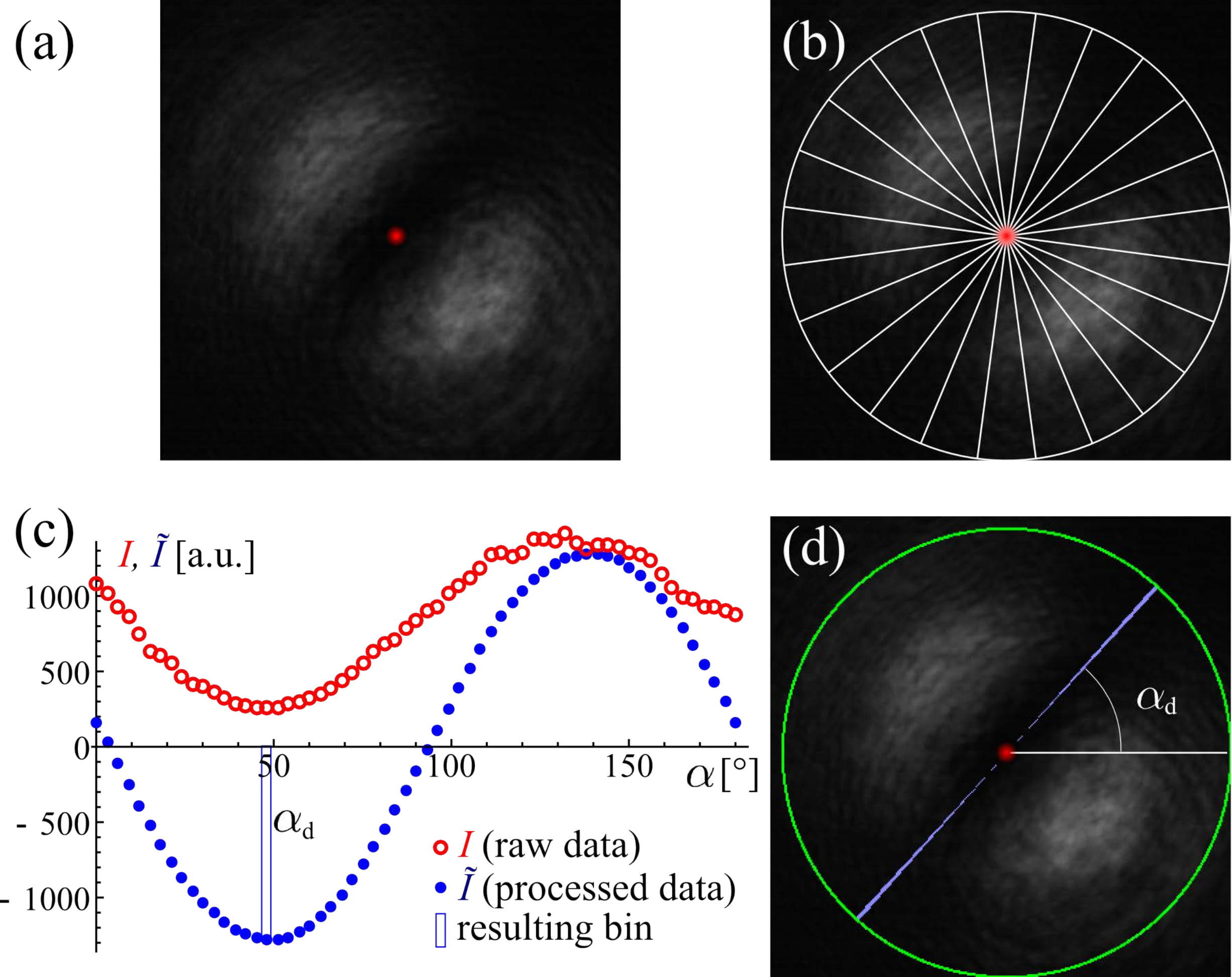}
\caption{Steps involved in the image analysis to determine the dark-line angle $\alphaD$. (a) Around the fitted center (indicated by a red spot), the image is split (b) in equally-weighted angular bins. On this figure, only 24 angular bins are shown for clarity, but $120$ bins are actually used in the real analysis. For each of them, the average intensity $I$ is computed. Half of these values (for $\alpha$ between 0 and \unit{180}{\degree}) are shown in the plot (c) as red open circles, as a function of the bin angle $\alpha$. The intensity values are processed as explained in the text to obtain the folded and averaged intensity values $\tilde{I}$. These are plotted as blue dots. In this example, the smallest value of $\tilde{I}$ was found in the \unit{48}{\degree} angle bin, which is marked by a blue box and corresponds to the angle \alphaD of the dark axis. In panel (d), the pixels belonging to this angular bin are colored in light blue to mark the dark axis, and the area of interest is indicated by the green circle. The original image size was $330\times330$ pixels. 
\label{Fig:referenceImageAnalysis}}
\end{figure}

\begin{figure}[b!!]
\includegraphics[width=0.96\linewidth]{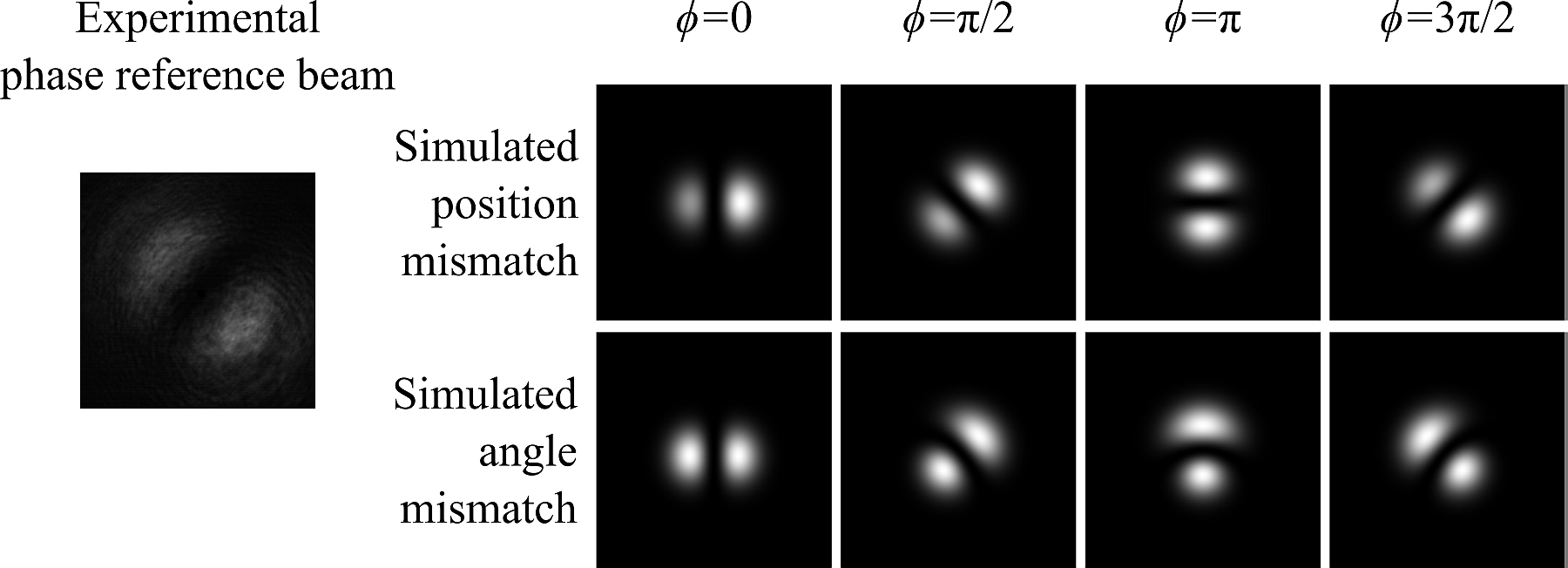}
\caption{Test of the robustness of the image analysis routine on a sample of test pictures. On the left of the figure is the experimental reference beam as it is seen on the camera for a certain value of $\varphi$. It is in fact the same image as in figure \ref{Fig:referenceImageAnalysis} for which our image analysis routine found $\varphi=48\degree$. On the right of the figure, we show a selection of simulated deformed beams. The beams were computed as the sum of two ideal LG beams with a relative phase $\phi$. In the top line, we simulated the effect of a small mismatch in the position of the center of the beams. In the bottom line, we simulated the effect of a small angle between the beams. Both defects yield a deformed HG mode in which one lobe is brighter than the other. This asymmetry is also present in the experimental reference beam on the left, but it is less pronounced than in the displayed simulations. When the simulated images were fed to the image analysis routine, the output was the correct $\phi$ value indicated above the pictures.}
\label{Fig:Test_routine}
\end{figure}

The following analysis of the individual images is illustrated in Fig.~\ref{Fig:referenceImageAnalysis}. The center is used as the origin of polar coordinates, while the radius of interest defines the area that will be analyzed. 
\begin{itemize}
\item This circular area is first divided into $N$ angular bins (``pie slices''), where $N$ has to be divisible by 8 so that there will be angle bins corresponding to each of the four \ketH, \ketD, \ketV and \ketA modes \footnote{The angle $\alpha \in \left[0,2\pi\right[$ is divided into $N$ angle bins, so the angle value of each bin is $\alpha_k = 2k\pi/N$, with $k \in [\![0,N-1]\!]$. The angle bin into which \alphaD must fall for projecting on mode \ketD has angle value $\alpha = \pi/4$, which implies $N=8k$.}. In our experiments, a typical value of $N$ was 120.
\item For each angular bin $k$, the average intensity $I(\alpha_k)$ is calculated. The first half of this data (\unit{0-180}{\degree}) is plotted in Fig.~\ref{Fig:referenceImageAnalysis}c as open red circles.
We could now fit a sinusoidal function to $I(\alpha_k)$ to determine \alphaD as its phase. However, since we have to process many images and since fits are computationally expensive, we choose the following straight-forward calculation instead.
\item Since we are only interested in an axis and not a direction, the angle bins are folded: The intensities of each two opposing bins, i.e. at \unit{180}{\degree} from each other, is added. This leaves us with only half the number of bins.
\item The dark line should be along the axis of least intensity, but also orthogonal to the axis with most intensity. To account for both conditions at once, we subtract from the intensity of each bin the intensity of the (unique) bin at \unit{90}{\degree} from it. These last two steps (folding and subtracting) also rectify slight image asymmetries that are always present (see explanations below and Fig.~\ref{Fig:Test_routine}).
\item To reduce the influence of remaining bin-to-bin noise, the data is smoothed: the processed intensity $\tilde{I}(\alpha_k)$ of bin $k$ is calculated as the average over a \unit{45}{\degree} wide sector centered around that bin. The last bin $N/2-1$ is considered being next to bin $0$, i.e. the angle is taken modulo \unit{180}{\degree}. The data $\tilde{I}$ is shown in Fig.~\ref{Fig:referenceImageAnalysis}c as blue dots.
\item The angle of the bin with the smallest $\tilde{I}$ is the angle \alphaD of the dark axis, providing the interferometer phase $\varphi$ via Eq.~\ref{eq:alphaDInterferometer}.
\end{itemize}
Since the alignment of the beams from the $L$ and $R$ paths is subjected to experimental imperfections, the image analysis routine has been tested against computer-generated images presenting various simulated defects.
For this, we numerically generated superpositions of Laguerre-Gaussian beams with either positional or angular misalignment.
Examples of such test images are given in Figure~\ref{Fig:Test_routine}.
The simulated beams have a more pronounced deformation than the experimental phase reference beam.
Yet, the image analysis routine yields the correct result for each of them, i.e., the algorithm finds the bin corresponding to the angle $\phi$ that was used to calculate the respective mode superposition.

\subsubsection{Timing and noise issues.}
\label{sect:Setup_TimingAndNoise}

The reference light is running backwards through the interferometer, but a tiny fraction of it is scattered towards the single-photon detectors at the outputs. To avoid any strong illumination of the detectors and to reduce potential noise in a first step, a very low power for the reference beam is required. The power chosen (\unit{\sim 2}{\nano\watt}) and the camera exposure time (\unit{\sim 100}{\milli\second}) finally resulted from a compromise between the reduction of this noise source and the recording of an image within a time shorter than the typical phase drift. 

Already at this low power level, the detector count rate due to the scattered reference light (\unit{10^6}{Hz}) is already close to saturation. During the signal measurements, we therefore interrupt the reference beam using acousto-optic modulators. This interruption is short compared to the duration of the experimental cycle, which is in turn much shorter than the exposure time of the reference camera, so this has no influence on the reference image acquisition. To illustrate this with numbers from our measurement presented later: the camera is taking a new reference image every \unit{125}{\milli\second}, with an exposure time of \unit{100}{\milli\second}. It is not synchronized to our experimental cycle, which interrupts the reference light every \unit{15}{\milli\second} for \unit{3}{\milli\second} to perform the APD measurements, leading effectively to 20\% less average intensity on the camera.

While the reference light was on, we switched off (gated) the single-photon detectors. Even in this configuration, we still noted an influence of the reference light on the background counts within the measurement window: when the reference light was completely blocked, the dark count rate was at about \unit{80}{\hertz}. With the reference light switched on as described above, the dark noise increased to \unit{200-250}{\hertz} within the measurement window. We excluded light leaking through the AOMs as the cause of the increased background count rates by additionally switching the reference light with mechanical shutters in a test. The dark count rate of the detector decreased over tens of milliseconds after turning off the relatively strong light exposure. Phenomenologically, the decay might be described by a stretched-exponential function\cite{Laherrere98Stretched}. This behavior has already been observed in other experiments, e.g.~\cite{Garcia13Fiber}, and it might be explained by delayed afterpulses of the avalanche photodiodes\cite{Enderlein05Using}.

\subsubsection{Polarization and wavelength of the reference beam.}
\label{PhaseReferenceWavelength}

We show now that it is highly desirable to use the same polarization and the same wavelength for the reference light as for the signal. 

First, the polarization-maintaining fibers and fibered beam-splitter, but also dielectric mirrors are birefringent, so the interferometer phase will be different for signal and reference if they have different polarizations. Furthermore, the birefringence changes with the mechanical stress of the fibers and with the temperature drifts, thus this difference will not stay constant. 

Second, if the two optical paths differ geometrically by $\Delta L$, two beams at different wavelengths will accumulate different interferometer phases. The variation of the measured interferometer phase around a certain wavelength $\lambda = c / \nu$ can be easily calculated as:
\begin{equation}\label{Eq:1_DeltaPhiGeometric}
(\dd\varphi / \dd\nu)  \approx12\degree / (\mathrm{cm\,GHz})\times\Delta L\;,
\end{equation}
where $\nu$ is the light frequency. 
So with a path difference $\Delta L$ on the order of a few centimeters, a difference by a few hundreds of MHz in the light providing the phase reference beam is already enough to change the inferred value of $\varphi$ by several tens of degrees.

Finally, dispersion can also play a significant role since a part of the interferometer is fibered. In a single-mode fiber, the change of the effective refractive index is dominated by the dispersion of the material\cite{Saleh1991Fundamentals_Ch8}. We can thus estimate the dispersion $\dd n / \dd \nu$ of our silica fibers to be on the order of \unit{10^{-3}/100}{\nano\meter} for our fibers \cite{Malitson65Interspecimen}. In a perfectly symmetrical situation, the first-order contribution of dispersion vanishes. However, even if the optical path lengths are precisely equal, there might be a difference in their composition in terms of free-space and fibered lengths. Let us call this difference in fiber length $\Delta L_\mathrm{fib}$. With this, we find:
\begin{equation}\label{EQ:2_DeltaPhiDispersion}
(\dd\varphi / \dd\nu) \approx -0.1 \degree/ (\mathrm{cm\,GHz}) \times \Delta L_\mathrm{fib}\;.
\end{equation}
This effect will be smaller than the previous one for typical configurations, but can still play a role if signal and reference are separated by several GHz.

If the frequencies of signal and reference are different, but stay constant, these two contributions lead, first, to a constant offset that could be determined, and second, to a different change of the phase when varying the path length. The latter difference is proportional to the relative wavelength difference and can thus in many cases be neglected for close wavelengths. As soon as the frequencies vary however, especially with respect to each other, the correlation between signal and reference interferometer phase will be lost. We therefore avoided these problems by using light from the same source as the one used for the signal state to be measured.

In this section, we have presented a detailed description of our OAM detection setup, and discussed the relevant issues relative to the phase-reference light. We now turn to the calibration procedure and provide the main figures of merit that have been measured thoroughly.

\section{Alignment procedure and calibration}
\label{sec:calib}

The detection setup presented here is very sensitive to the incoming beam position on the hologram dislocation and to its direction. The required fine tuning allows to calibrate and assess the performance of the setup, as shown in the following.

\subsection{Optimizing couplings and cross-talks}
The goal of the alignment procedure is twofold: 
an optimal coupling of the respective LG mode into the fiber of the respective branch and a strong rejection of all the other modes, as demanded by the mode selection requirement.
The alignment is performed using classical bright light fields aligned with the signal to be analyzed later, their path being defined by the same single-mode fiber. A spatial light modulator is inserted on the way and enables to send various spatial modes into the setup.  

In a first step, the position of the holograms is set by monitoring the intensity distribution with a camera placed behind the hologram. 
This distribution strongly depends on the relative position of the incoming beam and hologram center.
Sending in a \LG{+1} mode, we can thus optimize the observed intensity distribution to be close to a \G profile in the R path (Table \ref{Tab:Mode_conversion}). In the same way, the position of the L path hologram is optimized by sending in a \LG{-1} mode. Even small deviation of the hologram by a few micrometers become clearly visible. Then, the coupling into the fibers is optimized. \label{sect:Alignment:DiscussionCouplingEfficiencyAndRejection}Using two mirrors behind each of the holograms, we are able to adapt the mode exiting the hologram by maximizing the coupling efficiency up to $80\%$. 

The next stage consists in sending a \G mode and using the same mirrors to now minimize its coupling. The rejection pattern of the setup is more pronounced than the acceptance and thus allows a better approach to the optimal point. Finally, a random search in the region around the found optimum allows to do some fine tuning. Here, all 6 degrees of freedom (2 transversal positions of the hologram and 4 directions for the 2 mirrors) are slightly varied while switching between the coupled mode and the unwanted modes (such as \G). This way, the ratio $\eta_\mathrm{others}/\eta_\mathrm{LG}$ is minimized (where $\eta_{\Lambda}$ is the coupling efficiency of a mode $\Lambda$ into the single-mode fiber), while the coupling $\eta_\mathrm{LG}$ of the targeted mode is kept at or close to the maximum.

Each path is finally characterized by measuring the couplings for the different LG modes sent into the setup. In the R path for example, we record the transmission of the mode corresponding to that path (\LG{+1}), the two neighbors in $l$ number (\LG{+2} and $\LG{0}=\G$), and the mode corresponding to the other path (\LG{-1}). An average rejection of \unit{17}{\deci\bel} for the next neighbors ($2\%$ transmission) is typically obtained. For the suppression for the opposite mode (at $\Delta l=2$), a rejection of \unit{23}{\deci\bel} was obtained in the worst case, and up to \unit{37}{\deci\bel} in the best case. The typical value is \unit{25\pm3}{\deci\bel} \footnote{One of the holograms had a tiny scratch on its surface, leading to a lower efficiency.}. An example of detailed coupling figure of merit is given in Table~\ref{Tab:Calibration}.

\begin{table}[tbp]
\label{Tab:Mode_conversion}
{
\begin{tabular}[b]{|c|c c c|}
\cline{2-4}
\multicolumn{1}{c|}{}& \multicolumn{3}{c|}{Mode at input}\\
\multicolumn{1}{c|}{}&\LG{+1} & \G & \LG{-1}\rule{0pt}{3ex} \\
\hline
\multirow{5}{*}{\rotatebox{90}{R path}}& \multirow{5}{*}{\raisebox{-1.55ex}{\includegraphics[scale=0.065]{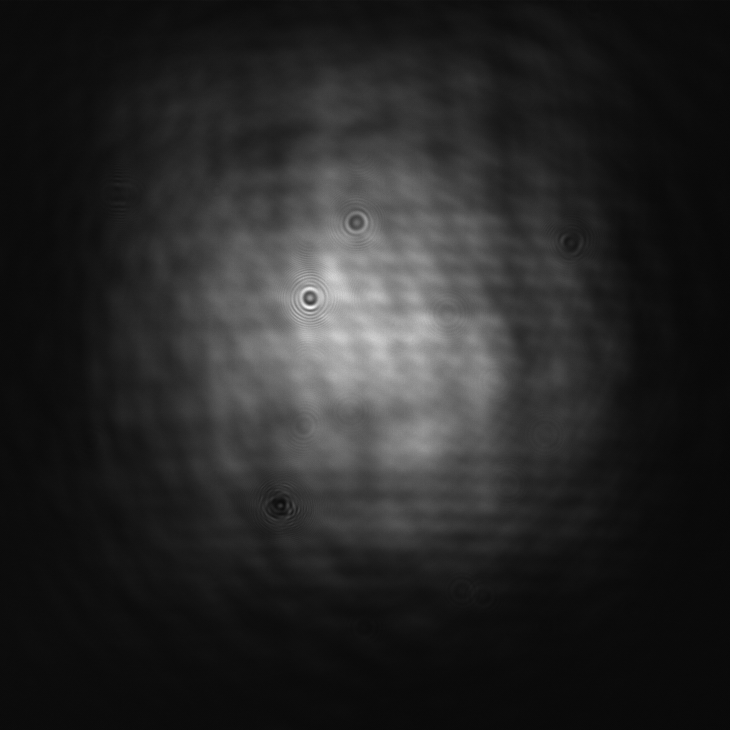}}} & \multirow{5}{*}{\raisebox{-1.55ex}{\includegraphics[scale=0.065]{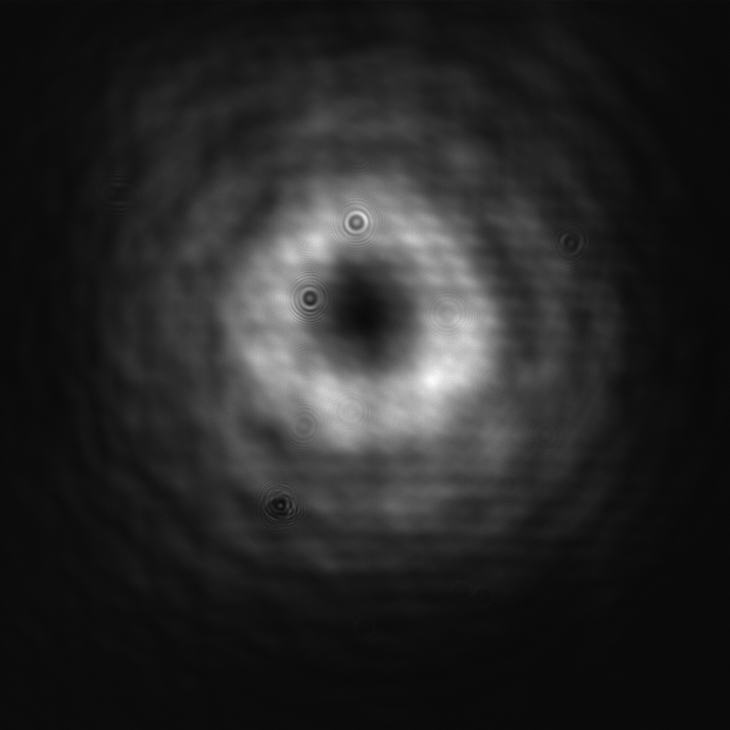}}} & \multirow{5}{*}{\raisebox{-1.55ex}{\includegraphics[scale=0.065]{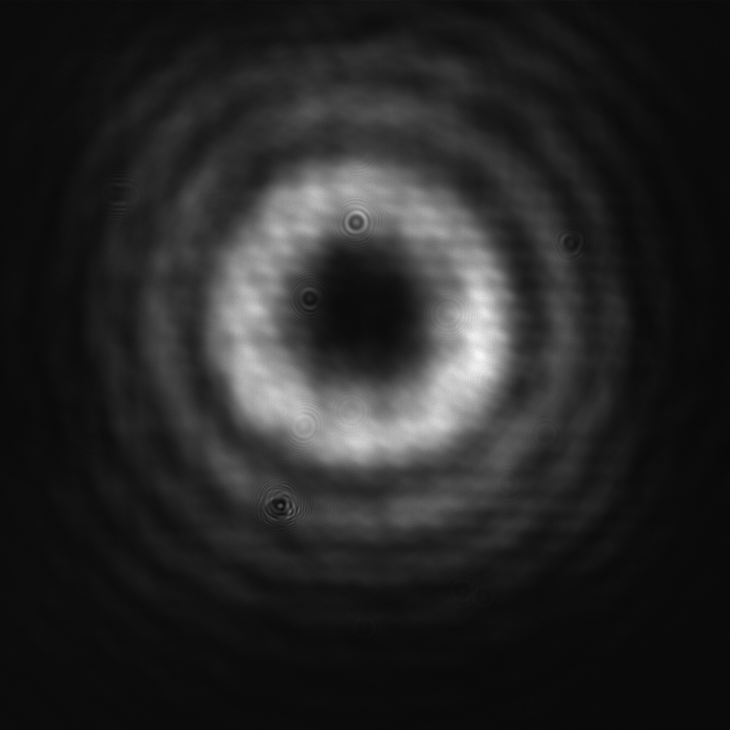}}} \\
& & & \\
& & & \\
& & & \\
& & & \\
\multirow{5}{*}{\rotatebox{90}{L path}}& \multirow{5}{*}{\raisebox{-1.55ex}{\includegraphics[scale=0.065]{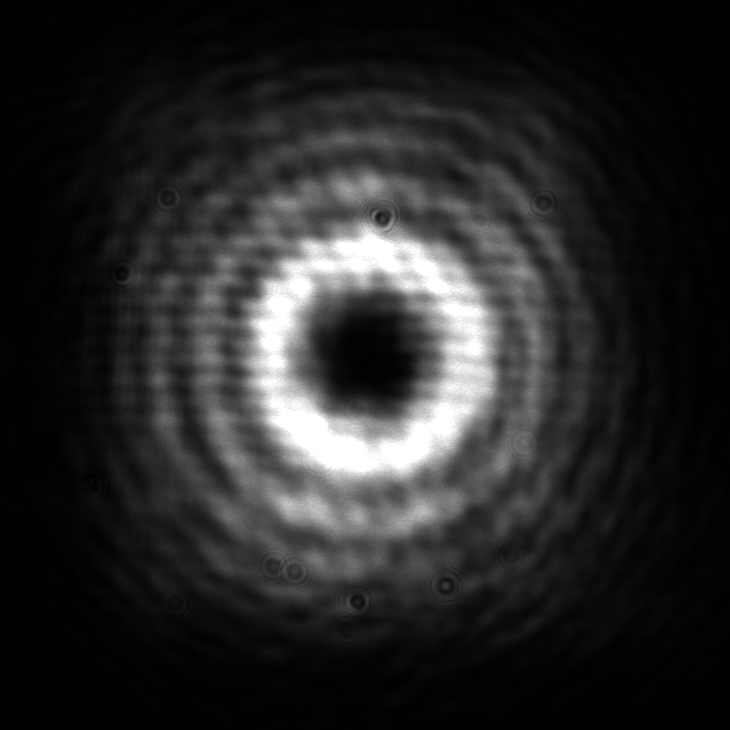}}} & \multirow{5}{*}{\raisebox{-1.55ex}{\includegraphics[scale=0.065]{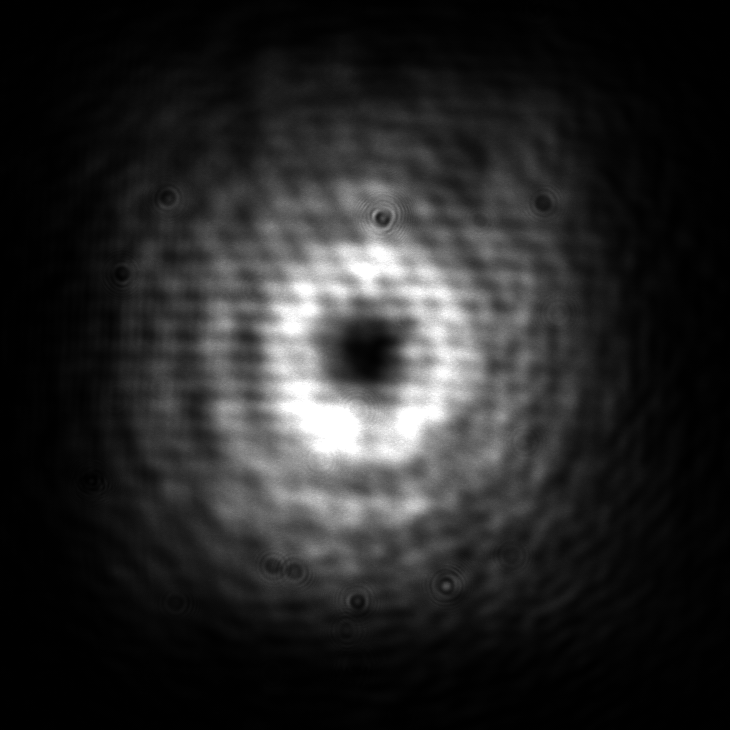}}} & \multirow{5}{*}{\raisebox{-1.55ex}{\includegraphics[scale=0.065]{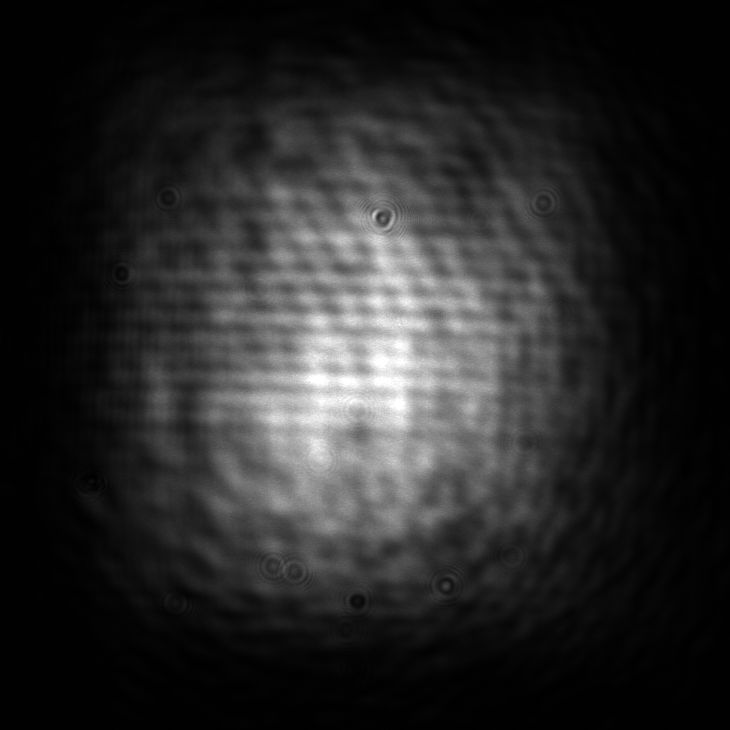}}} \\
& & & \\
& & & \\
& & & \\
& & & \\
\hline
\end{tabular}
}\hspace{2 mm}\vspace{2 mm}\label{Tab:Calibration}{
\begin{tabular}[b]{|l|r|r|}
\hline
\multicolumn{1}{|c|}{Mode}&\multicolumn{1}{c|}{R path}&\multicolumn{1}{c|}{L path}\\
\hline
\LG{-2}\rule{0pt}{3ex}&0.1 \%&2.8 \%\\
$\LG{-1}=L$&0.5 \%&\textbf{77.8} \%\\
\LG{0}&0.1 \%&1.7 \%\\
$\LG{+1}=R$&\textbf{82.3} \%&0.03\%\\
\LG{+2}&5.7 \%&0.04\%\\
\hline
\end{tabular}
}
\caption{Mode transformation and subsequent filtering. (a) Mode conversion performed by the holograms. After fine alignment of the hologram center, the impinging mode \LG{+1} (resp. \LG{-1}) is converted into a \G mode in the far field of the R path (resp. L path), while other modes are converted to higher $l$-valued modes with doughnut shapes. The goal of this alignment is to maximize the subsequent coupling of the desired mode in the optical fiber and to minimize the coupling of all other modes.
(b) Coupling efficiency in each path for various Laguerre-Gaussian modes at the input.}

\end{table}

\subsection{Calibration of the fringes}
\begin{figure}[t!]
\includegraphics[width=0.99\columnwidth]{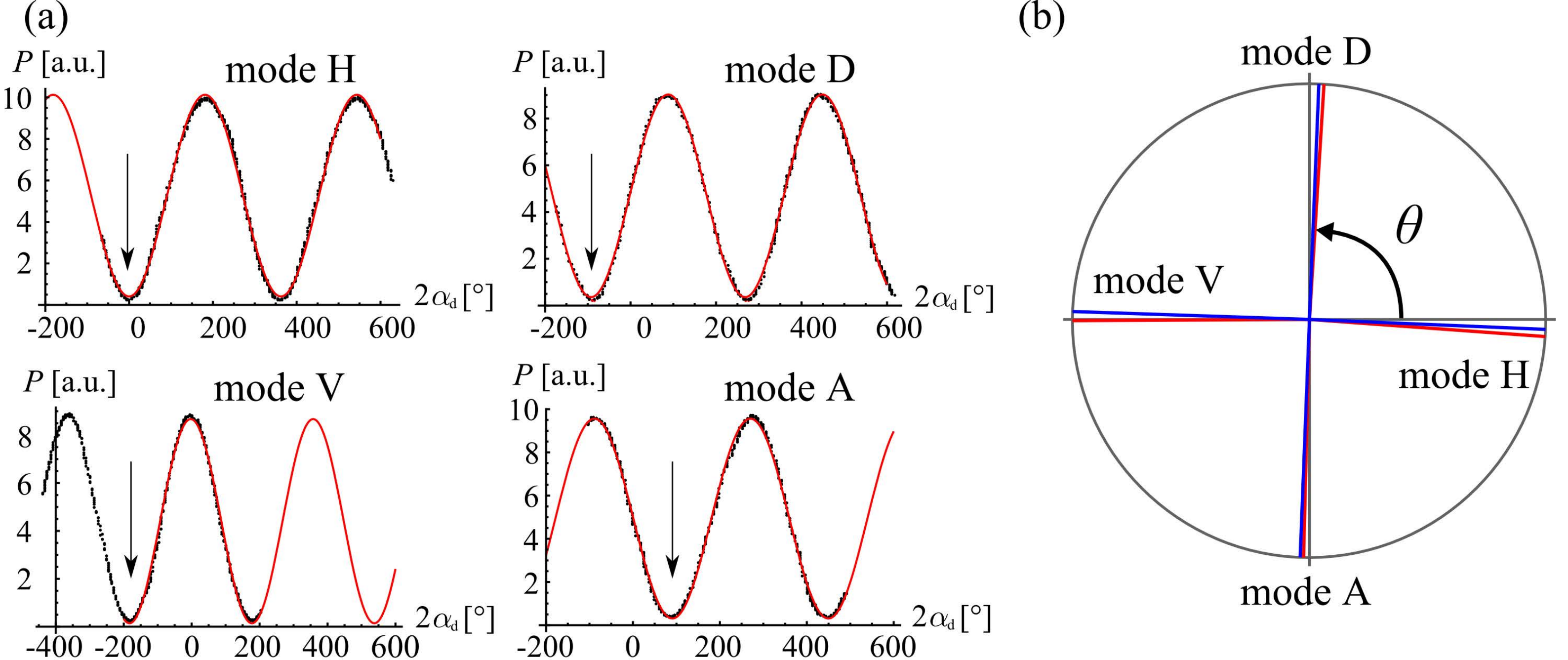}
\caption{\label{Fig:Interf_Calib}(a) Fringes measured during the calibration procedure for the various \HG{} modes at the input. Experimental points (black dots) corresponding to the power received at the output X are given together with a sinusoidal fit (solid red line). The horizontal axis corresponds to twice the angle $\alphaD$ of the dark line in the phase reference images. It is related to $\varphi$ by the relation $\varphi=2\alphaD+\pi$. As expected, the fringes shift as the various modes are sent. The position of the minima (indicated by arrows) gives the position where $\theta + \varphi \equiv \pi \pmod{2\pi}$. (b) From this measurement, we can locate the four \HG{} modes on the equator of the Bloch sphere (seen from top). The fringe positions measured in (a) are shown as red lines. The blue cross is obtained by averaging the deviations of the measured angles (red lines) from the respective theoretical angle (gray lines). In other words, it is the cross whose arms are the closest possible to the red lines while enforcing 90\degree{} between its arms. It shows an excellent agreement with the theoretical $\theta$ values ($0$, $\pi/2$, $\pi$ and $3\pi/2$ respectively for modes H, D, V and A). In this particular example,  the fringe visibility reaches $93 \pm 2\%$, and the inferred $\theta$ values were respectively -0.1\degree, 86.5\degree, 176.1\degree{} and 268.6\degree{} for the modes H, D, V and A, leading to an average deviation of $2.2\pm1.6\degree$. Let us note that, during the calibration measurement, the contamination with reference light is stronger than during the APD measurements for technical reasons. Therefore, the visibility is slightly reduced here in comparison with the data shown later.} \label{Fig:interf_calib}
\end{figure}

An additional characterization is performed by sending classical beams carrying the four Hermite-Gaussian modes H, V, D and A, as shown in Figure~\ref{Fig:Interf_Calib}. Theoretically, all four modes should lead to the same power of coupled light into both fibers. However, the power balance is not strictly mode-independent. Most of this imbalance can be explained by the imperfect mode filtering given in Table~\ref{Tab:Calibration}. In the R path for instance, $82\%$ transmission for mode \LG{+1} and $0.5\%$ for mode \LG{-1} make up for a $\pm 6\%$ difference in transmission between different HG modes. These mode-selective losses decrease the count rate and can additionally lead to a reduction of the fringe visibility, leading in turn to a decrease in the observed fidelity (see section \ref{subsec:imperfections}). 

Finally, the phase reference circuit is started and the interferometer phase slowly scanned. Sending the modes H, V, D and A, the respective fringes are measured and correlated to the interferometer phase obtained from the reference. 
These fringes allow to check first for a good fringe visibility (typical values above 93\% are obtained, mainly limited by the residual imbalance in the coupling of the HG modes and by the background noise) and second for the correct phase relation between signal and reference. 
Figure~\ref{Fig:Interf_Calib} illustrates this process. The power detected at the output X should exhibit a sinusoidal dependence in $\varphi = 2\alphaD+\pi$ (eq. \ref{Eq:alpha_dark_line}).
The condition for minimum power, i.e. $\varphi = \theta+\pi \leftrightarrow 2 \alphaD = \theta$ resulting from equation \ref{Eq:output}, allows to deduce the value of $\theta$ from the position of the fringes. We checked that this was indeed the case and found good agreement within $\pm 3 \degree$. 

The setup enables thus to accurately project an input state on various target modes. We now discuss the overall detection efficiency of the setup.

\subsection{Detection efficiency} 
\label{Subsec:Power_efficiency}

The overall loss of the device comes from various contributions:
\begin{itemize}
\item{Input-beam splitter and filtering: $\sim 50\%$. The combination of the $50/50$ beam splitter at the input and the subsequent mode selection causes a $50\%$ fraction of the signal to go ``into the wrong path'' whatever the input mode. This fraction is filtered out and lost. It does not account for the quality of the coupling into the mode selecting fibers.}
\item{Hologram diffraction and other optics: $\sim 80\%$ transmission.}
\item{Fiber coupling efficiency: $\sim 80\%$.} 
\item{Beam splitter for back-propagating the phase reference beam: $\sim 25\%$ losses. Larger splitting ratios, such as $90$:$10$ or even $99$:$1$ would reduce these losses, which are thus not intrinsic to our scheme.}
\end{itemize}

From all these parameters, one can extract the detection efficiency: $0.5\times 0.8 \times 0.8 \times 0.75 \sim 24\%\pm3\%$ (without including the quantum efficiency of the detectors). 

In our implementation, the detection efficiency for the HG modes was slightly lower than for the LG modes. This was due not to the detection setup but to the imperfection in the mode preparation. Indeed, the probe modes were generated by a reflection on a phase-only spatial light modulator. Yet, this method is limited by the intrinsic mode overlap between an ideal (target) mode and the sharply phase-modulated beam with gaussian envelope produced by the SLM. This overlap is lower for HG modes ($64\%$) than for LG modes ($78\%$) \cite{Morizur_10}. The rest of the power goes to higher-order $p>0$ modes that are not detected due to spatial filtering. In practice, this generates some additional mode-selective losses but cannot degrade the quality of the tomography (visibility and then fidelity), as they are the same for each mode in every pair of orthogonal modes.

\subsection{Impact of setup imperfections on the measurable fidelity}
\label{subsec:imperfections}

Experimental imperfections prevent from measuring unit fidelities even for ideal qubit states at the input. In this section, we discuss the upper limits on the fidelity values that can be measured.

The fidelity for HG states is especially sensitive to a reduced visibility $V$, setting a limit of $F_{\text{max}}=\sfrac{1}{2}(1+V)$. The reduced visibility can originate from a contamination with background noise, an imperfect mode filtering or a mode-dependent fiber-coupling. An imbalance of $\pm \Delta\eta$ in the fiber coupling of orthogonal HG modes will decrease the visibility of the fringe by approximately $(\Delta\eta)^2/2$. A $99\%$ fringe visibility, as achieved in some of the reported results without noise subtraction, leads to a maximum fidelity of $99.5\%$.

For LG modes at the entrance, no fringe should be present in the ideal case when scanning $\varphi$. Therefore, their fidelity is insensitive to the previous visibility reduction. However, a small relative leakage $\epsilon$ of mode L in path R (and vice versa) leads to a spurious fringe with a visibility $2\sqrt{\epsilon}$. This leakage translates into a maximum fidelity equal to  $1-\epsilon$. With a leakage of less than \unit{-25}{\deci\bel}, the resulting error on the fidelity is limited to a fraction of a percent.

Finally, a shift in the interferometer calibration by an angle $\Delta\theta$ will also decrease the maximum fidelity by an amount $(\Delta\theta)^2$. Even for a $\Delta\theta$ on the order of $5\degree$, the fidelity decreases thus by less than a percent. As can be seen from Figure~\ref{Fig:Interf_Calib}, this phase shift was controlled to better than $3\degree$ in our implementation.

\noindent In summary, this section has shown the reliability of our setup for OAM qubit measurements. Before moving on to some example of measurements performed with (approximate) single photons, let us recall three key parameters:
\begin{itemize}
\item{Cross-talk suppression (in LG basis): \unit{\sim 25}{\deci\bel},}
\item{Precision in phase measurement: better than \unit{3}{\degree},}
\item{Detection efficiency (excluding detectors): $24\% \pm 3\%$.}
\end{itemize}

\section{Example of quantum state tomography}
\label{sec:data}

We will now give some examples of density matrix reconstructions performed with our detection setup on weak coherent pulses in the single-photon regime. In this section, the setup is regarded as a black-box, yielding photon counts at the different outputs according to the interferometer configuration as described in Table~\ref{Tab:detected_modes}. Only data from output X is used for simplicity. 

The data presented here results from measurements on weak coherent states with a mean photon-number of $0.6$ photons per pulse. The spatial modes were imprinted by reflection on a spatial light modulator and were tuned to all of the six modes \Right, \Left, \ketH, \ketD, \ketV and \ketA. With the device in the HG bases configuration (i.e.\ measuring a fringe), three million measurements were made for each of the HG input modes, while one million measurements were performed for the LG input modes. In the configurations for measuring in the LG basis (i.e.\ blocked interferometer arm), one million measurements were performed for each of the two basis states, irrespective of the input state.

\begin{figure}[tbp]
\includegraphics[width=0.98\columnwidth]{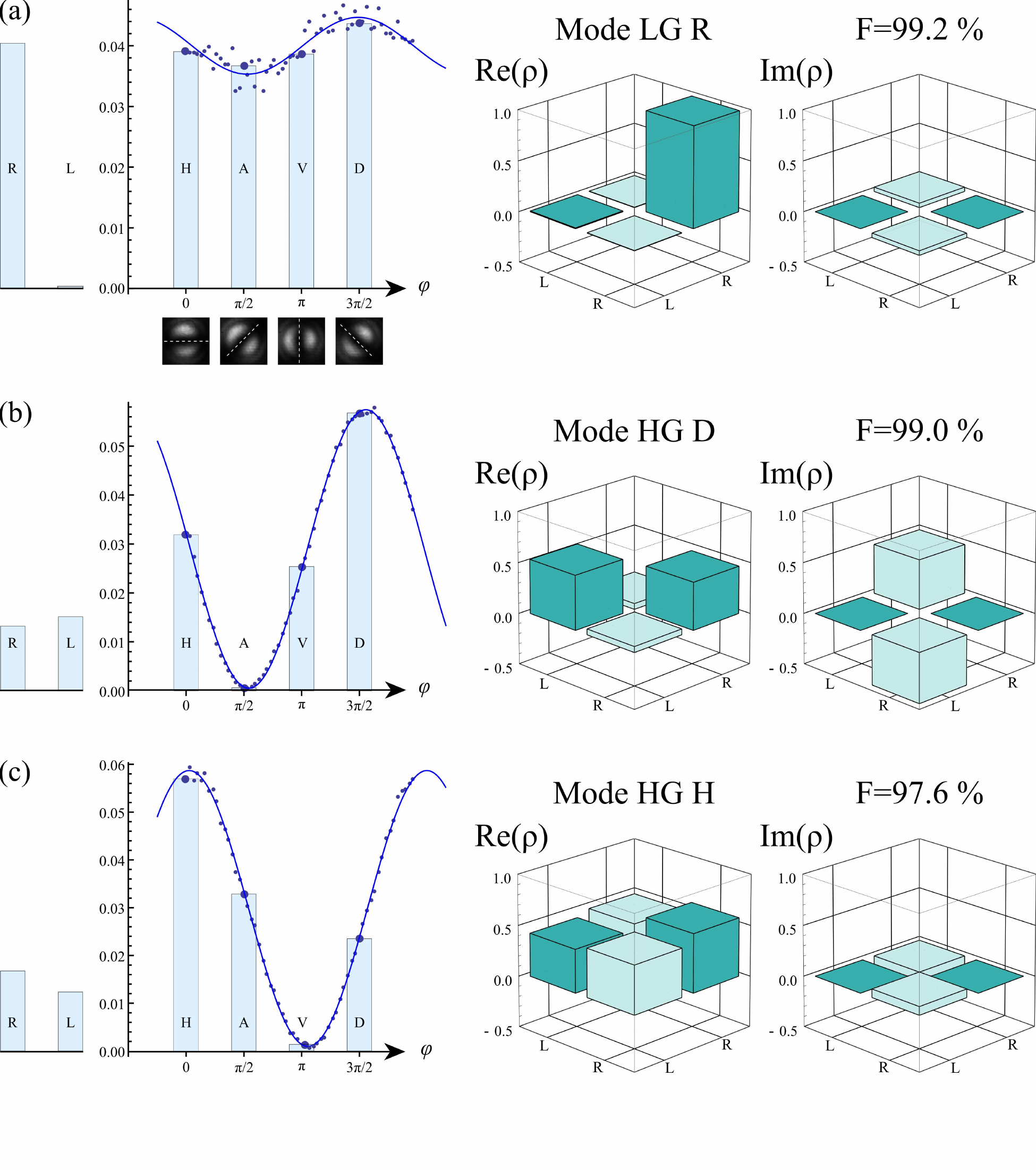}
\caption{\label{Fig:Fringe}Quantum state tomography of three different input states. The spatial mode is imprinted on attenuated coherent states with a mean photon-number of 0.6 per pulse. Panels (a), (b) and (c) correspond respectively to the \Right, \ketD and \ketH modes. The bar diagrams show the count rates (average number of detected photons per trial) recorded at the output X for various device configurations, corresponding to projections of the qubit over the states \Right, \Left, \ketH, \ketA, \ketV and \ketD. Typical images of the phase reference beam are shown below the axes in (a), corresponding to the values of $\varphi$ for the selected bins. The fringes superimposed on the bins (dots: experimental data, solid line: sinusoidal fit) show the variation of the count rates when the interferometer phase $\varphi$ is scanned. In (a), the spurious modulation in the count rate as a function of $\varphi$ comes from a small power leakage $\epsilon$ of the mode \Right into the Left path, as explained in section \ref{subsec:imperfections}. No background subtraction has been performed. The reconstructed density matrices are shown as well as the fidelity with the ideal state. The error bar on the fidelities is on the order of $\pm$1\%.
}
\end{figure}

From these measurements, the density matrix of each input state was reconstructed using the formulae given in section \ref{sec:qubits} : $S_1 = p_{\ket{0}} - p_{\ket{1}}$, $S_2 = p_{\ket{+}} - p_{\ket{-}}$ and $S_3 = p_{\ket{+\ii}} - p_{\ket{-\ii}}$, where $p_{\ket{\Psi}}$ is the probability to measure the qubit in the state $\ket{\Psi}$. In Table~\ref{Tab:Data_densitymatrix}, we give the density matrix parameters resulting from these calculations, and the fidelity to the ideal targeted state. In Figure~\ref{Fig:Fringe}, we give a more graphical view of these data for the three modes \Right and \ketD and \ketH. Fidelities from 96\% to 99\% are measured, mainly limited by the imperfect state preparation and the background noise, which has not been corrected here.

\begin{table}[b]
\centering
\begin{tabular}{|c|c|c|c|c|}
\hline
Mode & $S_1$ & $S_2$ & $S_3$& F \\ 
\hline \hline
\Right & 0.99 & 0.01 & -0.09 & 99.2\% \\ 
\Left & -0.97 & 0.0 & -0.14 & 98.3\% \\ 
\ketH & 0.15 & 0.95 & -0.17& 97.6\% \\ 
\ketD & -0.07 & 0.11 & 0.98 & 99.0\% \\
\ketV & 0.15 & -0.95 & -0.09 & 97.4\% \\
\ketA & 0.32 & -0.15 & -0.92 & 95.8\% \\
\hline
\end{tabular}
\caption{Stokes parameters and state fidelities extracted from the analysis of coherent pulses with a mean photon-number of $0.6$. In this particular example, the reconstruction of mode \ketA suffers from the coupling imbalance between paths R and L and from our imperfect mode preparation. The fidelity of the other modes is mostly limited by the background noise, which has not been subtracted here. The error bar on the fidelities is on the order of $\pm$1\%.
}
\label{Tab:Data_densitymatrix} 
\end{table}

For the sake of completeness, we finally give in the following some detailed values for this set of data:
\begin{itemize}
\item The measured average signal rate \nbarSig (i.e., the rate at the average height of the fringes) was 0.03~clicks/measurement.
\item The background noise originated from various sources. 
The intrinsic source is the APD dark noise of about \unit{200}{\hertz}, as discussed in Sect.~\ref{sect:Setup_TimingAndNoise}. The extrinsic sources are related to the environment in which our detection setup was used (a quantum memory experiment with at least one additional light beam during the measurements \cite{Nicolas_14}). With about \unit{1}{\kilo\hertz}, they constitute the major background contribution. An integration time of \unit{800}{\nano\second} per measurement gives thus a background rate of about $\nbarBg=10^{-3}$~background events per measurement ($\nbarBg/\nbarSig=3\%$).
\item While the total number of measurements is in the millions, the number of {\em events} in a single bin is much smaller, thus a significant Poissonian fluctuation occurs. The angular bins at different $\varphi$ values were only roughly equally covered by measurements, showing a standard deviation of up to 35\%. The average numbers of measurements per bin (statistical errors) were 560 (4\%) for the LG input states and 1400 (3\%) for the HG input states.
\end{itemize}

The interferometric scheme and the related calibrations enable thus to reconstruct the density matrix of qubits encoded in OAM states with fidelities close to unity. In the following, we discuss how to extend this approach to higher-dimensional space.

\section{Perspectives and possible extensions}
\label{sec:perspectives}

\begin{figure}[t!]
\includegraphics[width=0.99\columnwidth]{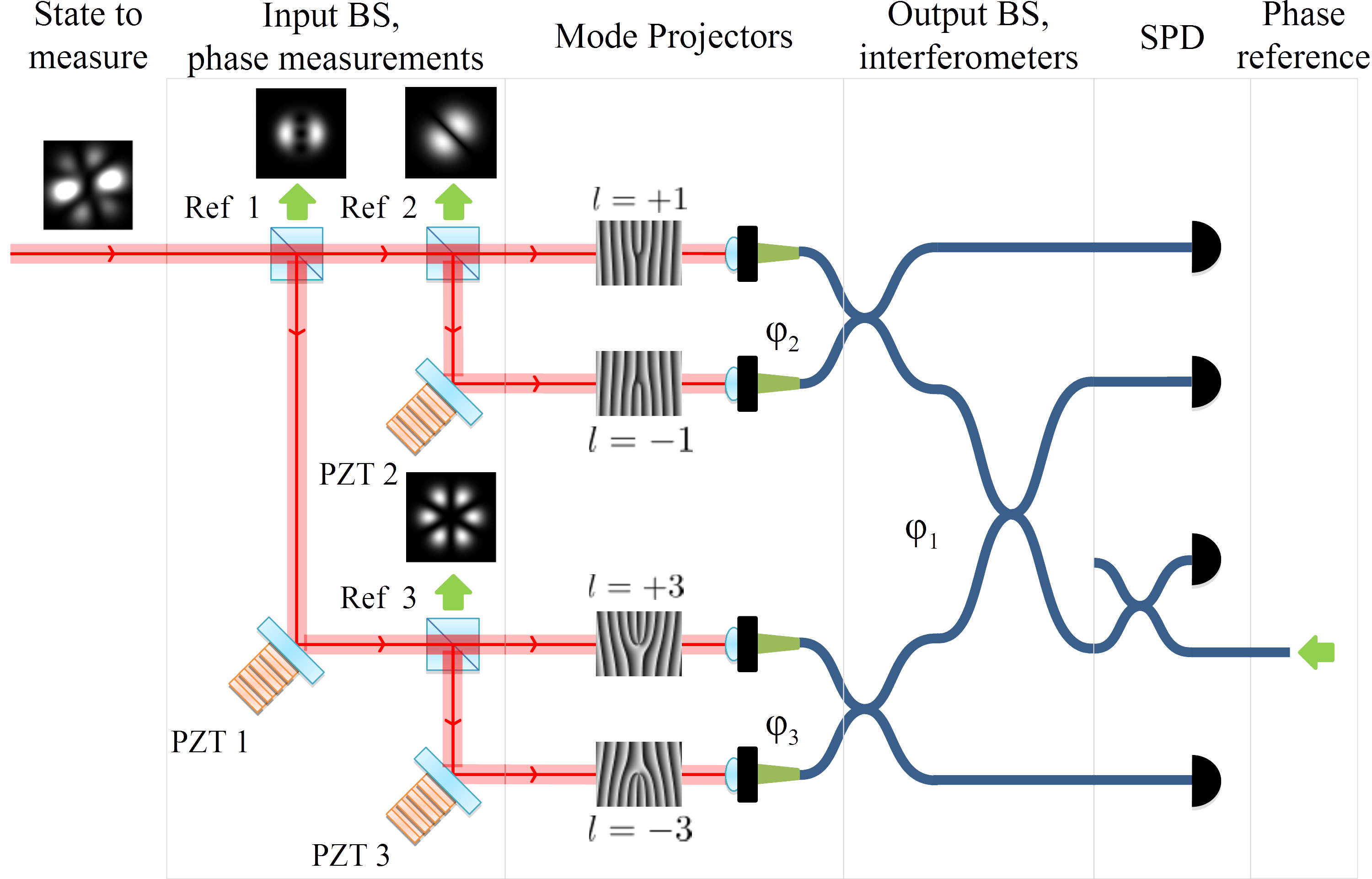}
\caption{Sketch of a similar setup for quantum state tomography in a four-dimensional Hilbert space spanned by $\ket{l=-3}$, $\ket{l=-1}$, $\ket{l=+1}$ and $\ket{l=+3}$. The mode to be measured is split between four paths by three cascaded beam-splitters. Each path contains a mode projector based on a fork hologram and a single-mode fiber. The transmitted light (corresponding to different projections on the OAM eigenstates) is brought to interfere in an array of beamsplitters and directed to single-photon counting modules. A phase reference beam (sketched by the green arrows) is injected backwards and detected at the three input beam-splitters. Ref. 1, 2 and 3 denote the outputs of the phase-reference beam allowing to measure the phases $\varphi_{1,2,3}$ respectively. Expected intensity patterns are displayed. In order to get the Ref. 1 intensity pattern (which allows a simple measurement of $\varphi_1$), the $l=+1$ and $l=-3$ paths must be blocked, e.g. using fast switching mechanical shutters. The mode displayed at the input, i.e. $\ket{l=+1}+\ket{l=-1}-\ket{l=+3}-\ii\ket{l=-3}$, is shown for illustrative purpose.}
\label{Fig:FourModesDevice}
\end{figure}

To reach higher-dimensional space, one can introduce additional beam-splitters after the first input one. Figure \ref{Fig:FourModesDevice} shows the example of an extended setup to perform quantum state tomography in a four-dimensional Hilbert space spanned by the modes $\LG{-3}$, $\LG{-1}$, $\LG{+1}$ and $\LG{+3}$. 
In each of the subsequent paths, a mode projector on a different OAM value is inserted. 
Keeping an OAM difference $\Delta l = 2$ between different modes ensures a better mode filtering.
Fibers at the end of the mode projectors are connected to a series of cascading fiber beam-splitters, creating an array of nested Mach-Zehnder interferometers.
Alternatively, these nested interferometers could be engraved in photonic circuitry \cite{Shadbolt_11,Dai_12,Spagnolo_13}, which would also provide greater simplicity and better phase stability.
Similarly to the two-dimensional setup, a phase reference beam is sent backwards.
The various unused output ports allow the imaging of the phase reference beams, and the determination of all the relevant phases.
The previously described image-analysis routine (\ref{pythonroutine}) can be directly used to compute the relevant phases between pairs of modes, given the fact that the phase reference is timed in order to image only two mode superpositions.

\begin{figure}[t!]
\includegraphics[width=0.98\columnwidth]{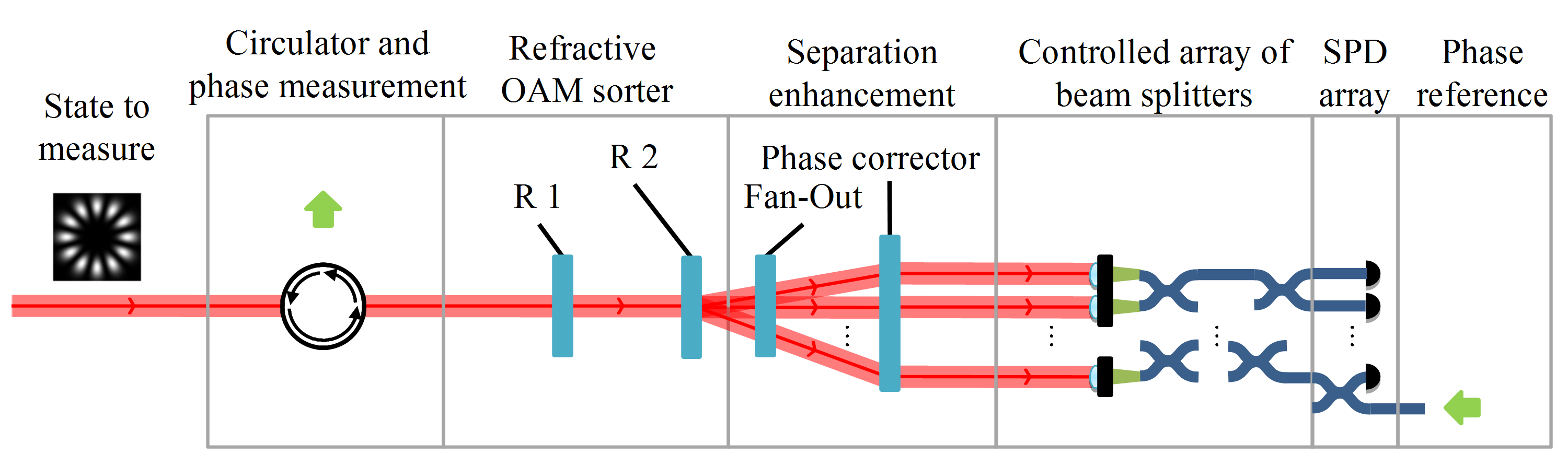}
\caption{A possible alternative implementation for extending the device to higher-dimensional Hilbert spaces, using mode sorting refractive optics R1 and R2 as developed in \cite{Lavery_12,Mirhosseini_13}. The input beam-splitters and subsequent holograms can be replaced by the combination of a circulator (to separate the signal and the phase reference beam) and an OAM mode sorter made of two refractive elements R1 and R2. The reformatter R1 performs a log-polar to Cartesian coordinate mapping and the phase corrector R2 corrects for the different propagation lengths from R1 to R2. These two elements transform co-propagating LG beams with different $l$ values into approximate \G beams with a propagation direction tilted by an angle proportional to $l$. The fan-out separation enhancer is a specific periodic phase-only hologram that increases the angular separation between the modes by decreasing the residual mode-overlap. After being separated and losing their specific spatial shapes, the different OAM components can be brought to interfere in an array of beam-splitters forming a mesh of interferometers with controllable phases, then directed towards single-photon counting modules. A good timing of the phase reference should be ensured with appropriated shutters inserted in the interferometer array in order to record all the relevant phase differences. The superposition $\ket{l=+6}+\ket{l=-6}$ is shown at the input for illustrative purpose.}
\label{Fig:Mode_sorter_extension}
\end{figure}

One more series of beam splitters would allow quantum tomography in an eight-dimensional space, but it would become challenging. 
Indeed, the addition of more beam-splitters will result in degrading the count rates exponentially in the number of beam splitters (linearly in the number of detected modes).
Also the phase measurement and/or stabilization can become a more serious issue if the dimension of the Hilbert space increases. 

Another way to reach higher-dimensional spaces would be to take advantage of the recently developed OAM mode-sorting techniques \cite{Lavery_12, Sullivan_12, Mirhosseini_13,Malik_14} where OAM states are converted into transverse momentum states.
As shown in Figure~\ref{Fig:Mode_sorter_extension}, the input beam-splitter can indeed be replaced by an OAM mode sorter made out of refractive elements as described in \cite{Lavery_12} with a separation enhancer \cite{Mirhosseini_13}. This combination would perform both the separation and mode conversion in the same time, thus largely improving the detection efficiency and versatility.
Indeed, the detection efficiency would remain (almost) constant as the number of detected modes increases.
The reduction in the required number of optical elements may also provide a better phase stability. Table \ref{Tab:possible_extensions} summarizes the performances one can foresee using some realistic parameter estimations extracted from \cite{Malik_14}.

\begin{table}[t!]
\centering
\begin{tabular}{|c|c|c|c|}
\hline
\bf Technique:&\multicolumn{2}{c|}{beamsplitter cascade} & mode sorting \\
\hline
\bf Device:& Current & 3BS & OAM sorter \\ 
\hline 
\hline
Hilbert space & \multirow{2}{*}{2} & \multirow{2}{*}{4} & \multirow{2}{*}{15} \\ 
dimension & & & \\
\hline 
Losses & 75\% & 88\% & 40\%\\ 
\hline 
Crosstalk  & & & \\
Suppression & $>\unit{27}{\deci\bel}$ & $>\unit{27}{\deci\bel}$ & $>\unit{30}{\deci\bel}$ \\ 
($\Delta l = 2$) & & & \\
\hline 
\end{tabular} 
\caption{Expected performances of two possible extensions of the current device (``Current'') to higher dimensional Hilbert spaces. The current device enters into the category ``Beamsplitter cascade'', although the cascade is only one step deep.   The device ``3BS'' refers to the setup shown in Fig.~\ref{Fig:FourModesDevice} and extends the current device to a two-step cascade of beamsplitters. ``OAM sorter'' refers to the device shown in Fig.~\ref{Fig:Mode_sorter_extension} exploiting a mode sorter scheme. Detection efficiencies does not take into account the intrinsic efficiency of the single-photon counters.}
\label{Tab:possible_extensions} 
\end{table}

\section{Conclusion}\label{conclusion}
In conclusion, we have presented and quantitatively characterized a setup enabling the quantum state tomography of a photonic qubit encoded in the orbital angular momentum degree of freedom. This interferometric setup performs the projection of an input state over different mutually unbiased bases and enables consequently the reconstruction of the full density matrix. Various noise and imperfection sources have been discussed and the benchmarks of the experimental implementation have been given. We have finally proposed some extensions of our scheme to higher-dimensional Hilbert spaces and have shown that the combination of the presented technique with the recent and elegant mode-sorting methods may be applied to experiments requiring OAM qudit tomography. 

\begin{acknowledgments}
The authors thank A. Zeilinger and R. Fickler for providing fork holograms and M.J. Padgett and D. Tasca for their assistance with SLM. This work is supported by the ERA-Net CHIST-ERA (QScale), the ERA-Net.RUS (Nanoquint) and by the European Research Council (ERC Starting Grant HybridNet). This work has benefited from discussions within the CAPES-COFECUB project Ph 740-12. A.N. gratefully acknowledges support from the Direction G\'en\'erale de l'Armement (DGA). J.L. is a member of the Institut Universitaire de France.
\end{acknowledgments}

\bibliography{99_biblio}

\end{document}